\documentclass[a4paper,twocolumn,11pt,accepted=2021-03-15]{quantumarticle}
\pdfoutput=1
\usepackage[utf8]{inputenc}
\usepackage[english]{babel}
\usepackage[T1]{fontenc}
\usepackage{amsmath}
\usepackage{hyperref}

\usepackage{tikz}
\usepackage{lipsum}

\usepackage{latexsym}
\usepackage{graphics}
\usepackage{graphicx}
\usepackage{epsfig}
\usepackage{color}
\usepackage{braket}
\usepackage{lipsum}
\usepackage{siunitx}
\usepackage{comment}
\usepackage{physics}
\usepackage[numbers,sort&compress]{natbib}

\newcommand{\mathcall}[0]{}
\newcommand{\red}[1]{#1}

\begin{document}
	
	\title{Entangled resource for interfacing single- and dual-rail optical qubits}
	\author{David Drahi}
	\email{david.drahi@physics.ox.ac.uk}
	\affiliation{Clarendon Laboratory, Department of Physics, University of Oxford, Oxford OX1 3PU, UK}
	
	\author{Demid V. Sychev}
	\affiliation{Russian Quantum Center, 100 Novaya St., Skolkovo, Moscow 143025}
	\affiliation{Moscow State Pedagogical University, M. Pirogovskaya Street 29, Moscow 119991, Russia}
	
	\author{Khurram K. Pirov}
	\affiliation{Moscow Institute of Physics and Technology, 141700 Dolgoprudny}
	
	\author{Ekaterina A. Sazhina}
	\affiliation{Russian Quantum Center, 100 Novaya St., Skolkovo, Moscow 143025}
	\affiliation{Moscow Institute of Physics and Technology, 141700 Dolgoprudny}
	
	\author{Valeriy A. Novikov}
	\affiliation{Niels Bohr Institute, University of Copenhagen, DK-2100 Copenhagen, Denmark}
	
	\author{Ian A. Walmsley}
	\affiliation{Clarendon Laboratory, Department of Physics, University of Oxford, Oxford OX1 3PU, UK}
	\affiliation{Imperial College London, Exhibition Road, London, SW7 2AZ, UK}
	
	\author{A. I. Lvovsky}
	\email{Alex.Lvovsky@physics.ox.ac.uk}
	\affiliation{Clarendon Laboratory, Department of Physics, University of Oxford, Oxford OX1 3PU, UK}
	\affiliation{Russian Quantum Center, 100 Novaya St., Skolkovo, Moscow 143025}
	\affiliation{P. N. Lebedev Physics Institute, Leninskiy prospect 53, Moscow 119991, Russia}
	\homepage{http://quantech.group}

\begin{abstract}
	Today's most widely used method of encoding quantum information in optical qubits is the dual-rail basis, often carried out through the polarisation of a single photon. On the other hand, many stationary carriers of quantum information --- such as atoms --- couple to light via the single-rail encoding in which the qubit is encoded in the number of photons. As such, interconversion between the two encodings is paramount in order to achieve cohesive quantum networks. In this paper, we demonstrate this by generating an entangled resource between the two encodings and using it to teleport a dual-rail qubit onto its single-rail counterpart. 
	This work completes the set of tools necessary for the interconversion between the three primary encodings of a qubit in the optical field: single-rail, dual-rail and continuous-variable.
\end{abstract}

\section{Introduction} 
With the development of quantum technology, it is becoming clear that different physical systems are optimal for various aspects of quantum information processing. For example, superconducting circuits and trapped ions are well-suited for the implementation of quantum logic gates for computation and simulation; spin ensembles for quantum memories; natural and artificial atoms for precise sensing. A comprehensive quantum network should enable reliable exchange of quantum information among all these systems \red{\cite{Awshalom2019}}. The primary agent of such exchange is light, as it is the only physical system able to carry quantum information over large distances. Technologies of quantum coupling between light and stationary carriers of quantum information, such as superconducting cavity modes \cite{Andrews2014}, spin ensembles in solids and atomic gases \cite{Lvovsky2009}, optomechanical devices \cite{Aspelmeyer2014} and others are being developed. 
	

However, these technologies must address an important challenge before their broad deployment becomes possible. In many promising quantum settings --- for example, excitons in quantum dots, single atoms and superconducting qubits --- the natural basis for encoding the qubit consists of two energy eigenstates. When coupled to light, such a qubit is naturally converted into the so-called single-rail qubit: the encoding in which the vacuum $\ket{0}$ and single-photon $\ket{1}$ states correspond to the two logical states. Single-rail qubits are however notoriously inconvenient when it comes to quantum information processing and communication by means of light. This is because single-qubit operations are difficult in this encoding \cite{Berry2006}. Additionally, the information carried by a single-rail qubit can be easily distorted by optical loss, which results in $\ket{1}$ becoming $\ket{0}$.

A much more robust way of encoding the optical qubit is dual-rail, in which the logical states consist of the photon occupying one of the two orthogonal optical modes. In this way, the photon is present in any valid state of the qubit, thereby providing an easy way to identify when the qubit has been lost. The two ``rails" can be, for example, different time or frequency bins, or the orthogonal polarisations, horizontal $\ket{H}$ and vertical $\ket{V}$. The latter approach --- utilised in this work --- is particularly common, as it allows easy realisation of single-qubit operations by means of polarisation rotators. 

In principle, the dual-rail light qubit can be treated as a pair of single-rail qubits carried by each polarisation mode, as we can write $\ket{H}=\ket{1_{H}}\ket{0_{V}}$ and $\ket{V}=\ket{0_{H}} \ket{1_{V}}$. That is, a general dual-rail qubit can be split via a polarising beam splitter into two spatial modes, each of which would \red{need} to be individually coupled to a stationary system, as demonstrated in e.g. Ref.~\cite{Vernaz2018} and references therein. However, this approach necessarily demands a doubling of the number of components in the network. Further, it would still require a method for single-to-dual rail interconversion within the stationary system, which would need to be specific for each such system. It therefore appears more practical to develop a method for such interconversion only for light. 

\begin{figure*}[t!]
	\includegraphics[width=18cm]{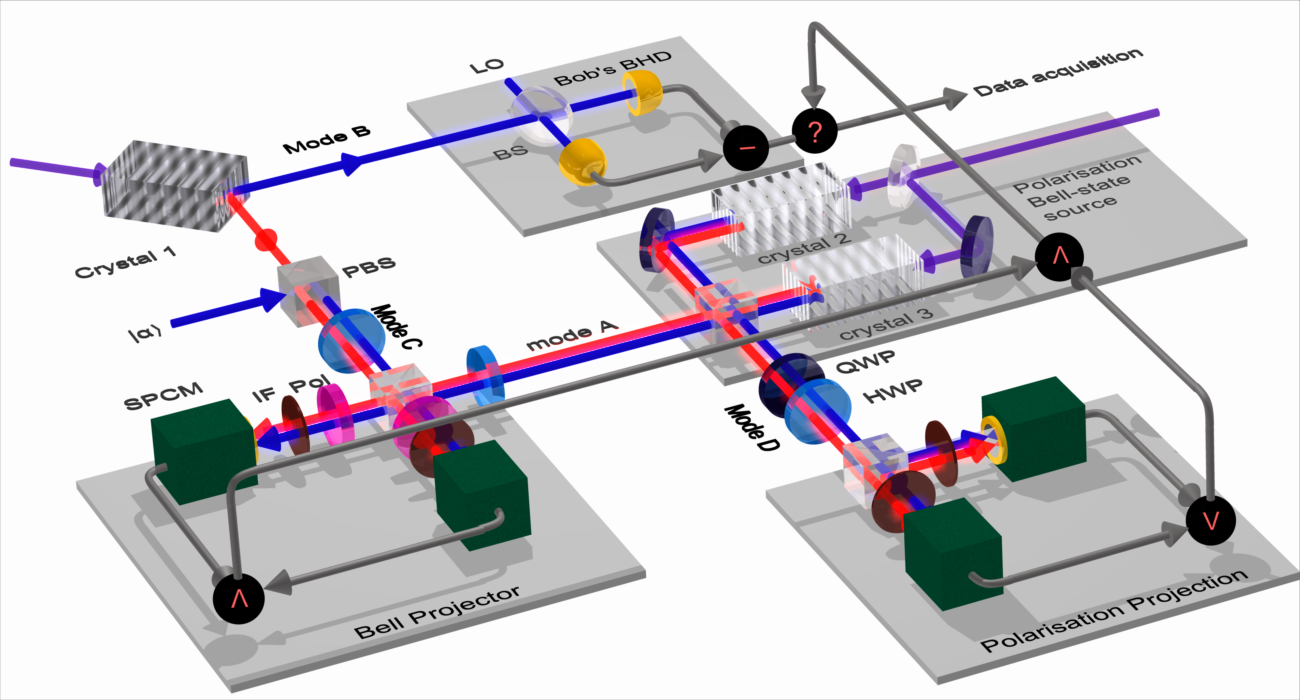}
	\caption{Experimental setup. To prepare the entangled resource \eqref{eq:Resource1} in spatial modes C (dual-rail) and B (single-rail), one combines a photon from a pair generated in crystal 1 with a coherent state $\ket{\alpha}$ on a PBS. Alice's source state $\ket{\chi}_{A}$ is obtained \red{in mode A} by applying a polarisation projection in mode D to a Bell pair $\ket{\Psi}_{AD}$ produced by crystals 2 and 3. HWP: half-wave plate; QWP: quarter-wave plate; Pol: polariser; \red{IF: interference filter;} (P)BS: (polarising) beam splitter; SPCM: single-photon counting module; BHD: balanced homodyne detector; LO: local oscillator. The red (blue) lines correspond to horizontal (vertical) polarisation, parallel red and blue lines to dual-rail qubits, grey lines to electronic signals. \red{Black spheres mark arithmetic and logical operations: conjunction ($\land$),  disjunction ($\lor$), subtraction ($-$), and conditioning (?).}}
	\label{fig:setup}
\end{figure*}

To date, there existed methods for preparing entangled states that connect single- and dual-rail qubits with ``continuous-variable" \red{(``cat")} qubits carried by coherent states of opposite phases \cite{Morin2014,Jeong2014,Sychev2018}. In principle, one could use these resources to convert between single-and dual-rail encodings through an intermediate step of continuous-variable encoding. However, this approach is quite cumbersome and susceptible to error. It would be much more desirable to develop a direct method for such interconversion.

This is the goal of our paper. We propose and implement a technique to prepare an entangled resource of the form 
\begin{equation}
\ket{\aleph}=a \ket{H} \ket{1} + b \ket{V} \ket {0} \,.
\label{Psi}
\end{equation}
We show that this resource can be used for the interconversion between the two bases via quantum teleportation \cite{Bennett1993} from a qubit carried by a photon's polarisation (which we associate with the fictitious observer Alice) onto the single-rail encoding (received by observer Bob). Specifically, we prepared all 6 primary basis states of a dual-rail discrete variable qubit $a\ket{H} + b\ket{V}$ and teleported them onto their single-rail counterparts $a\ket{0} + b\ket{1}$. In this aspect, our experiment achieves the goal pursued in theoretical proposals \cite{ralph2005adaptive,Fiuracek2017}, albeit with a different method which is more general, more experimentally accessible and less vulnerable to inefficiencies. 

\begin{figure*}[t!]
	\includegraphics[width=18cm]{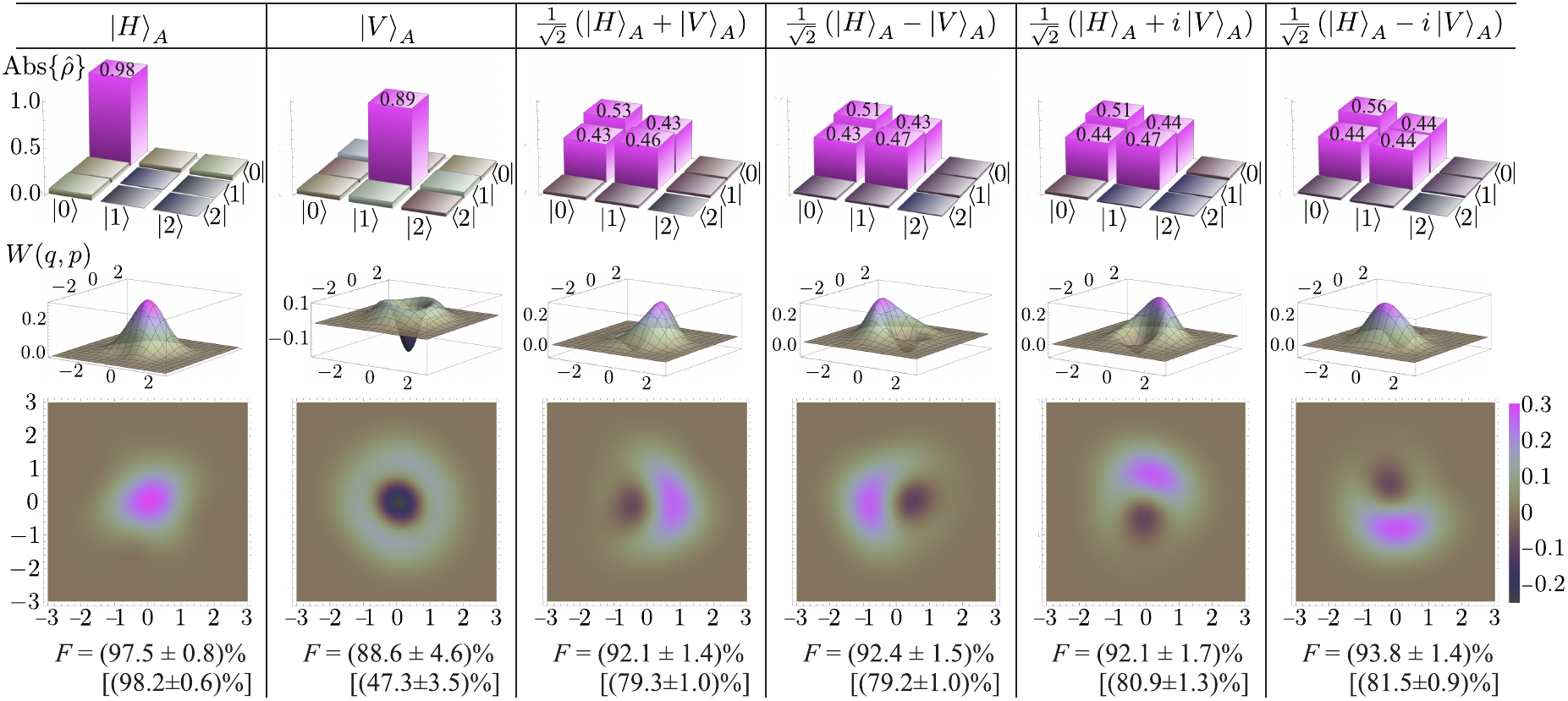}
	\caption{Teleportation experiment results. The top line shows the dual-rail source states prepared by Alice. The resulting teleported single-rail qubits\red{, reconstructed with a 50\% efficiency correction,} are shown below. From top to bottom, we display the absolute value of the reconstructed density matrices $\mathrm{Abs}\{\hat{\rho}\}$ in the Fock basis, the Wigner functions $W\left(q,p\right)$ in phase space and their surface plots. Fidelities $F$ calculated with respect to the expected pure states are given withal\red{, both with and without [in square brackets] efficiency correction}. The fidelity uncertainties include both the statistical error of the state tomography and the error arising in the efficiency-corrected state reconstruction due to the efficiency drift (see Appendix). }\label{fig:Teleportation}
\end{figure*}

\section{Theory}To produce the entangled resource state, we begin from non-degenerate spontaneous parametric down-conversion (SPDC) in crystal 1 as shown in Fig.~\ref{fig:setup}. This produces the state $\ket{\psi}_{\mathrm{cr1},CB}=\ket{0_{H}}_{C}\ket{0_{V}}_{B} + \gamma_{1} \ket{1_{H}}_{C}\ket{1_{V}}_{B} + \mathcal{O}(\gamma_{1}^{2})$ in the vertical and horizontal polarisations of \red{spatial} modes B and C, respectively, with $\gamma_{1}$ being the SPDC amplitude. We then use a polarising beam splitter (PBS) to inject a weak coherent state $\ket{\alpha_{V}}_{C}=\ket{0_{V}}_{C} + \alpha \ket{1_{V}}_{C} + \mathcal{O}(\alpha^{2})$ into the vertical polarisation of mode C. The collective state describing these two \red{spatial} modes can be written as the tensor product 
\begin{equation}
\begin{split}
\ket{\Omega}_{CB}&\equiv\ket{\alpha_{V}}_{C}\otimes\ket{\psi}_{\mathrm{cr1},CB}\approx \ket{0_{H}0_{V}}_{C}\ket{0_{V}}_{B} \\
&+ \gamma_{1}\ket{1_{H}0_{V}}_{C}\ket{1_{V}}_{B} + \alpha\ket{0_{H}1_{V}}_{C}\ket{0_{V}}_{B}\\
&=\ket{0}_{C}\ket{0}_{B} + \gamma_{1}\ket{H}_{C}\ket{1}_{B} + \alpha\ket{V}_{C}\ket{0}_{B}\,,
\end{split}
\label{eq:Resource1}
\end{equation}
up to the first order, where we chose the dual- and single-rail notations in mode C and the vertical polarisation of mode B, respectively, in the last line. 

Although this state is separable, entanglement of the form \eqref{Psi} is present in its last two terms corresponding to a photon present in mode C. Intuitively, this reflects the ambiguous situation in which this photon could originate from either crystal 1 or the weak coherent state $\ket{\alpha}$. If such photon came from crystal 1 ($\ket{\alpha}$), then Bob receives a single photon $\ket{1}$ (vacuum $\ket{0})$ in the vertical polarisation of mode B. 

Theoretically, this entanglement could be recovered by means of a non-demolition (for example parity \cite{gerry2005quantum}), polarisation-insensitive detector in mode C that would project that mode onto the single-photon state. Nonetheless, even without such a detector, the state $\ket{\Omega}_{CB}$ can be used post-selectively to perform the teleportation. 

To generate the input state for the teleportation, Alice starts by producing an entangled state $\ket{\Psi}_{AD}=\ket{0}_{A}\ket{0}_{D} + \gamma_{2,3}\left(\ket{H}_{A}\ket{V}_{D} + \ket{V}_{A}\ket{H}_{D}\right)$ between spatial modes A and D. This is achieved by means of SPDC crystals 2 and 3 (see Fig.~\ref{fig:setup}). Then, a polarisation projection of $\ket{\Psi}_{AD}$ performed in mode D by one of the two single-photon counting modules (SPCMs) prepares a heralded photon, carrying a dual-rail qubit of the form $\ket{\chi}_{A}=a\ket{H}_{A} + b\ket{V}_{A}$, in mode A. To teleport this state, Alice sends it into a Bell state projector $\bra{\Psi^{+}}_{AC}=\frac{1}{\sqrt{2}}\left(\bra{H}_{A}\bra{V}_{C} + \bra{V}_{A}\bra{H}_{C}\right)$, combining it with the part of the resource state $\ket{\Omega}_{CB}$ in mode C. Upon a successful application of the Bell projector --- characterised by a coincidence click from the two SPCMs in Fig.~\ref{fig:setup} --- Bob finally obtains the following single-rail teleported state in his mode B
\begin{equation}
\begin{split}
\ket{\varphi}_{B}&=\bra{\Psi^{+}}_{AC}\ket{\chi}_{A}\ket{\Omega}_{CB}\\
&=\frac{1}{\sqrt{2}}\left(a\,\alpha\ket{0}_{B} + b\,\gamma_{1}\ket{1}_{B}\right)\,.
\end{split}
\label{eq:TeleportedState}
\end{equation}
For faithful teleportation, we match the coherent state's amplitude to that of SPDC from crystal 1 so that $\alpha=\gamma_1$.

The same scheme, but without a measurement in mode D, can produce a freely-propagating single-dual entangled resource \eqref{Psi}. This happens thanks to entanglement swapping when $\ket{\Omega}_{CB}\ket{\Psi}_{AD}$ is projected onto the Bell state $\ket{\Psi^{+}}_{AC}$ in \red{spatial} modes A and C. This, however, would require a source of Bell states in modes A and D operating in a heralded \cite{Barz2010, Wagenknecht2010} or deterministic \cite{Muller2014} fashion. In the absence of such a source, we can still show the viability of this procedure by reconstructing the resulting state in modes B and D post-selectively \cite{Sychev2018} and assessing its \textit{a posteriori} entanglement. 

\section{Experiment} 
The master laser is a pulsed Ti:Sapphire (Coherent Mira 900D) with a wavelength of $780\,\si{\nano\meter}$, mean output power of $1.3\,\si{\watt}$, repetition rate of $R_{L}=76\,\si{\mega\hertz}$ and pulse width of $1.6\,\si{\pico\second}$. This laser provides the weak coherent state $\ket\alpha$ and the local oscillator for the homodyne detection; its second harmonic also pumps the three parametric down-conversion crystals used across the setup.

We perform frequency doubling in a lithium triborate crystal with an efficiency of $\sim30\%$. After further mode filtering, $50\,\si{\milli\watt}$ and $5\,\si{\milli\watt}$ of the frequency doubled wave pump crystal 1 and both crystals 2 and 3, respectively. These are periodically poled potassium titanyl phosphate crystals (PPKTP) operating in a type II spectrally and spatially degenerate, but polarisation non-degenerate configuration. The down-conversion amplitudes were set, for the reason explained below, to unequal values of $\gamma_{1}\approx0.20$ and $\gamma_{2,3}\approx0.054$. 

To generate Alice's entangled state $\ket{\Psi}_{AD}$, we used the Mach-Zehnder interferometer architecture introduced in \cite{Kwiat_HVVH_Source}, where the outputs of two SPDC crystals interfere on a PBS. Due to different optical path lengths in the Mach-Zehnder interferometer, photons from crystals 2 and 3 experience a phase difference $\Delta\phi_{\gamma_{2,3}}$ which is introduced in $\ket{\Psi}_{AD}$. We use a piezoelectric transducer along with an interferometric feedback loop to ensure $\Delta\phi_{\gamma_{2,3}}=0$ at all times. The quality of the entangled state $\ket{\Psi}_{AD}$ has been tested by applying a polarisation projection in mode D and measuring the probability of a coincidence click between a pair of SPCMs in modes A and D as a function of the angle of a half-wave-plate (HWP) set in mode A. The resulting probabilities depend sinusoidally on that angle, with measured visibilities of $99\%$, $97\%$ and $95\%$ for projections of $\ket{\Psi}_{AD}$ onto the canonical, diagonal and circular polarisation bases in mode D, respectively \red{\cite{thesis}.} . 


The central element of the Bell state projection apparatus is a PBS, which directs the photons of different polarisations into the same spatial mode, thereby preventing coincidence clicks from the input states $\ket{\Psi^\pm}_{AC}$. The photons in the two PBS output modes are filtered by polarisers set at $\pm\frac{\pi}{4}$ prior to detection, so that the state $\ket{\Phi^{+}}_{AC}$ is not transmitted. Therefore, a coincidence click projects the PBS input onto $\ket{\Phi^{-}}_{AC}$ \cite{Pan_Bell_Projector, Kim2003, AlexBook}. However, because both input modes of the Bell projector are rotated, using HWPs, by $\frac{\pi}{4}$ before entering the PBS, the state to which the Bell projector responds is $\ket{\Psi^{+}}_{AC}$. 

Our implementation of the projector is advantageous with respect to the version that makes use of a sole 50:50 beam splitter \cite{Braunstein_Bell_Projector}. First, it reduces the rate of false positive Bell detection events due to two photons present in the same input mode, being oblivious to those of them in which these two photons have orthogonal polarisations. Second, observing polarisation projections of $\ket{\Omega}_{CB}$ in mode C allows us to monitor in real-time the phase in channel B necessary for quantum state reconstruction of Bob's teleported state (see Appendix). Third, it obviates the need to align the regular beam splitter to an exact 50\% reflectivity. 

Prior to the teleportation experiment, we tested the projector's alignment by preparing a heralded photon in mode A in the vertical polarisation ($\ket{\chi}_{A}=\ket{V}_{A}$) and observing its Hong-Ou-Mandel interference with the weak correspondingly polarised coherent state $\ket{\alpha_{V}}_{C}$ in mode C. A visibility of $98\%$ was observed.

\section{Results}
To reconstruct the teleported states in Bob's channel, we employed a balanced homodyne detector (BHD) \cite{Kumar2012, Masalov2017}. A total of 2000 quadratures --- along with their associated phases --- were acquired for each teleported state. \red{The states have then been reconstructed} using the maximum-likelihood algorithm \cite{Lvovsky_MaxLikelihood} \red{both with and without correcting for the experimental efficiency. To evaluate the correction factor, we performed homodyne tomography of the heralded Fock state in mode B \cite{lvovsky2001quantum}, conditioned on the event in one of the SPCMs in modes A and C. We found the detected state to be a mixture of the single-photon and vacuum states with the single-photon fraction $\eta = (50\pm2.5)\%$ (see Appendix), and we use this value for efficiency correction. This choice allows us to separate the imperfections of the novel features of the quantum communication protocol developed here from the technical imperfections arising from inefficient preparation and detection of the single photon.}

The teleportation has been implemented for six input states: $\ket{V}$, $\ket{H}$, $\frac{1}{\sqrt{2}}\left(\ket{H}\pm\ket{V}\right)$ and $\frac{1}{\sqrt{2}}\left(\ket{H}\pm i \ket{V}\right)$. The results of the experiment are shown in Fig.~\ref{fig:Teleportation}. The output states show an average fidelity of $(92.8\pm2.8)\%$ with respect to the expected pure states in Eq.~(\ref{eq:TeleportedState})\footnote{The fidelity between two states with density operators $\hat\rho$ and $\hat\sigma$ is defined as $\mathcall{F}=\left({\mathrm Tr}\sqrt{\sqrt{\hat\sigma}\hat\rho\sqrt{\hat\sigma}}\right)^2$.}. For clarity, we only show the density matrices up to 2 photons in Fig.~\ref{fig:Teleportation}, while the states used to compute the fidelity have been in fact reconstructed up to 4 photons. For the diagonal and circular inputs, the orthogonality between the resulting qubits, $\ket{\pm}_{B}=\frac{1}{\sqrt{2}}\left(\ket{0}_{B} \pm \ket{1}_{B} \right)$ and $\ket{\pm i}_{B} = \frac{1}{\sqrt{2}}\left(\ket{0}_{B} \pm i \ket{1}_{B} \right)$ manifests itself in the Wigner functions oriented in opposite directions. 

The probability of teleportation events scales as $p_{\mathrm{good}}\sim\eta_{d}^3\left|\gamma_{1}\right|^2\left|\gamma_{2,3}\right|^2$, where $\eta_{d}\approx 3\%$ is the total quantum efficiency associated with single-photon detection. The measured rate of these triple events was $R_{T}\sim0.16\,\si{\hertz}$ (see Appendix). The main source of error explaining the deviation of the experimentally observed teleported states from those predicted by Eq.~(\ref{eq:TeleportedState}) comes from the false positive Bell state projections \cite{braunstein1998posteriori, pan2003experimental}: a coincidence click can be caused by both photons arriving from the same input mode, A or C, rather than one from each mode. The probability of such events for the two photons coming from mode A scales as $p_{\mathrm{bad,A}}\sim2\eta_{d}^3\left|\gamma_{2,3}\right|^4$, and for the photons from mode C as $p_{\mathrm{bad,C}}\sim2\eta_{d}^3\left|\gamma_{1}\right|^4\left|\gamma_{2,3}\right|^2$. These values are different because one of the photons arriving from mode A is always heralded by a click in mode D. To minimise the ratio of these events with respect to the ``good" ones while keeping the latter at a reasonable rate, we set $\left|\gamma_{2,3}\right|\sim\left|\gamma_{1}\right|^2$. This results in the false positives contributing $~\sim16$\% to all triple coincidences. Most of these events lead to the admixture of the vacuum state to Bob's output. Because the theoretically expected output state is expected to contain, on average, one-half of the vacuum state, the effect of this admixture on the fidelity is $\sim8\%$, consistent with our observations. 

If the teleported state reconstruction is performed without efficiency correction, the resulting average fidelity with respect to the expected pure states in Eq.~(\ref{eq:TeleportedState}) is reduced to $(77.7\pm1.5)\%$, thus remaining at the state-of-the-art level \cite{Pirandola2015NatPhotReview}.

\begin{figure}[h!]
	\includegraphics[width=\columnwidth]{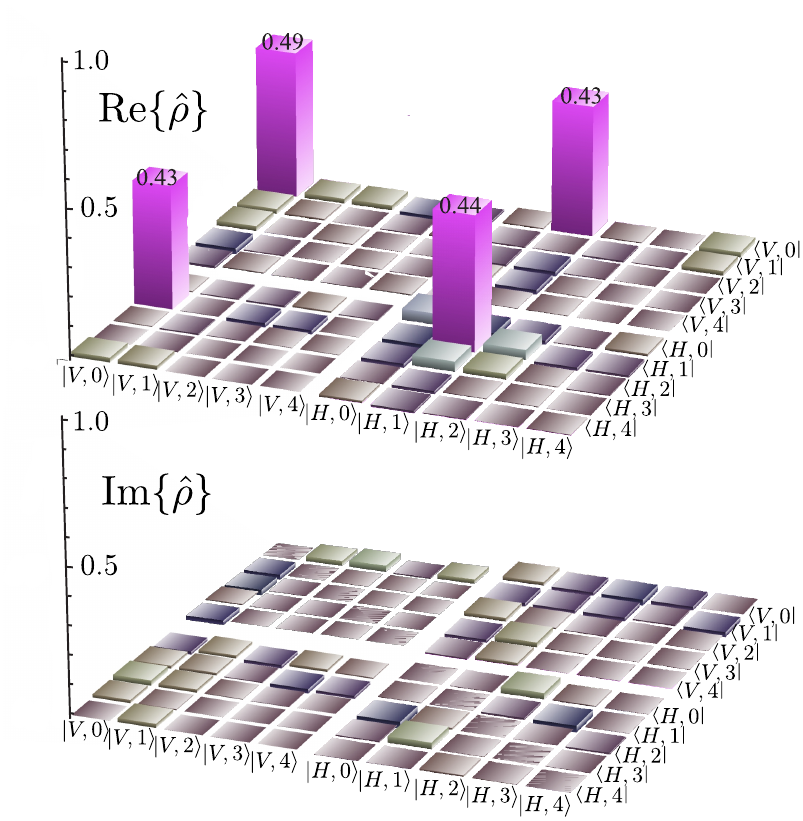}
	\caption{Entanglement swapping result. Density matrix $\hat\rho$ of the post-selectively reconstructed state obtained through entanglement swapping in  modes B and D after the Bell state measurement in modes A and C. It is presented in the basis $\{\ket{H,n},\ket{V,n}\}$, where the polarisation and Fock states label the dual-rail mode D and single-rail mode B, respectively.}
	\label{fig:swapping}
\end{figure}

The teleportation outputs for the six inputs allow us to post-selectively reconstruct the bipartite state that is produced in modes B and D due to entanglement swapping after the Bell measurement in modes A and C \cite{Sychev2018}. Comparing the result displayed in Fig.~\ref{fig:swapping} with the expected maximally entangled state of Eq.~(\ref{Psi}), we find a fidelity of $(89.7\pm2.7)\%$ $\left[(66.8\pm1.3)\%\right]$ with [without] applying efficiency correction to the maximum likelihood algorithm, thus certifying the \textit{a posteriori} entanglement of this state since the classical limit of 1/2 \cite{sackett2000experimental} is beaten. 

\section{Summary} 
In summary, we have proposed and experimentally demonstrated a scheme to prepare a fully entangled resource state $\ket\aleph$ of dual- and single-rail optical qubits. This state enables exchange of quantum information between these two encodings by way of teleportation, as we show here by converting qubits from dual- to single-rail encoding. 
For inverse conversion, a Bell measurement in the single-rail qubit basis would need to be constructed. Projecting onto single-rail qubit Bell states $\frac1{\sqrt2}(\ket 0\ket 1\pm\ket 1\ket 0)$ can be realised by overlapping the two modes on a symmetric beam splitter. This would transform these states into $\ket0\ket 1$ and $\ket 1\ket 0$, which can then be detected via efficient photon counters \cite{babichev2003quantum}.

\red{In our experiment, the state $\ket\aleph$  is implemented postselectively.} If our scheme is used in combination with heralded or deterministic sources of entangled photons (which is within the reach of modern experimental technology \cite{Barz2010,Wagenknecht2010,Muller2014}), it would produce $\ket\aleph$ in a heralded fashion through entanglement swapping. By storing this state in a quantum memory and retrieving it on-demand, it could be used in practical quantum networking to enable efficient exchange of quantum information between stationary carriers of different nature by means of light. 

There are three primary ways to encode a qubit in the optical field: single-rail, dual-rail and continuous-variable. While previous research \cite{Morin2014,Jeong2014,Sychev2018} established techniques to connect the two discrete-variable encodings with the continuous-variable one, the present work completes the triad to enable interconversion among all three encodings. \red{Note, however, that this triad does not yet include the Gottesman-Kitaev-Preskill continuous-variable qubits, which are considered promising for optical quantum computing \cite{GKP}. This challenge has to be addressed by future research.}

\red{The code for maximum-likelihood state reconstruction and fidelity calculation, as well as the raw quadrature data and the reconstructed density matrices are available in an online repository \url{https://github.com/trekut/H1V0}.}

\begin{acknowledgments}
	The work was supported by the Russian Foundation for Basic Research (Grant No. 18-37-20033) and the Russian Science Foundation (Grant No. 19-71-10092). 
\end{acknowledgments}


\bibliographystyle{plainnat}
\bibliography{H1V0citations1}

\onecolumn\newpage
\appendix

\section{Efficiency estimation} 
We measure the quantum efficiency $\eta$ from the quantum state tomography of a single Fock state \cite{lvovsky2001quantum}. The deviation of $\eta$ from unity arises from three main factors: losses, mode matching between the signal and local oscillator (LO) and the quantum efficiency of the homodyne detector's photodiodes at the laser's wavelength. Their estimated values are $0.80$, $0.81$ and $0.86$, respectively\red{, with a product of $0.56$}. Additionally, during the data acquisition runs, the high pump power impinging onto crystal 1 systematically caused the SPDC mode's quality to degrade over time as a result of grey tracking in the PPKTP crystals. Averaging over the duration of an acquisition batch, we have $\eta = 50\%$ with a drift of $\pm2.5$\%.

\section{Phase reconstruction}
Precise determination of the teleported state's phase with respect to the local oscillator is essential for its reconstruction via homodyne tomography. Taking into account the difference between the phase $\phi_{\alpha}$ of the state $\ket{\alpha}$ and the phase $\phi_{\gamma_{1}}$ of the photons output by crystal 1, Bob's teleported state (Eq.~(3) in the main text) becomes
\begin{equation}
\begin{split}
\ket{\varphi}_{B}&=\frac{1}{\sqrt{2}}\left(a\,\alpha\ket{0}_{B} + b\,\gamma_{1} e^{-i\left(\phi_{\gamma_{1}}-\phi_{\alpha}\right)} \ket{1}_{B}\right)\,.
\end{split}
\label{eq:TeleportedState_Phase}
\end{equation}

To evaluate this phase, we use one arm of the Bell projector to perform a polarisation projection of the resource state $\ket{\Omega}_{CB}$ in mode C onto the diagonal basis. When a single click is recorded, we get the state
\begin{equation}
\ket{\delta}_{B}\equiv\left(\frac{\bra{H}_{C} + \bra{V}_{C}}{\sqrt{2}}\right)\ket{\Omega}_{CB}=\frac{1}{\sqrt{2}}\left(\,\alpha\ket{0}_{B} + \,\gamma_{1} e^{-i\left(\phi_{\gamma_{1}}-\phi_{\alpha} \right)} \ket{1}_{B}\right)\,,
\label{eq:RMP_Phase}
\end{equation}
whose phase equals that of the teleported state in Eq.~(\ref{eq:TeleportedState_Phase}).  

The rate of these single clicks is five orders of magnitude higher than the successful triple coincidence clicks required for teleportation. Therefore, by continuously monitoring the mean value of the quadrature observable $\hat{X}$ in Bob's channel conditioned on these frequent single clicks, we can \red{estimate} the phase of $\ket{\delta}_{B}$ with a 1.4\% accuracy. We also note that, because of the the long acquisition time, the phases of the acquired quadrature sample sets are virtually uniformly distributed over the range from 0 to $2\pi$.

\section{Photon count rates} 
Photons are measured by SPCMs from Excelitas. The typical count rates measured in both arms of the Bell projector were $R_{\alpha} = R_{\gamma_{1}} = 22\,\si{\kilo\hertz}$ for $\ket{\alpha}$ and crystal 1. For crystals 2 and 3, these rates were $R_{\gamma_{2,3}} = 1.7\,\si{\kilo\hertz}$. These numbers reflect the loss factor of 4 introduced by the pairs of HWPs and polarisers in the Bell projector. The coincidence rate of the photons originating from crystals 2 and 3, measured between one of the polarisation analysis SPCMs in mode D and one of the Bell projector SPCMs, was, on average, $R_{\gamma_{2,3}}^{\mathrm{c.c}}= 51\,\si{\hertz}$. From these rates, we estimated a total photon detection efficiency of $\eta_{d} = \frac{R_{\gamma_{2,3}}^{\mathrm{c.c}}}{R_{\gamma_{2,3}}} \approx 3.0\%$. This value is rather small due to the presence of narrowband ($0.2\,\si{\nano\meter}$) filters in front of each SPCM, as well as the unavoidable optical losses present in the experiment. From here, we estimate $\gamma_{1}=\left(\frac{R_{\gamma_{1}}}{R_{L}}\frac{1}{\eta_{d}}\right)^{1/2}\approx0.20$ and $\gamma_{2,3}=\left(\frac{R_{\gamma_{2,3}}}{R_{L}}\frac{1}{\eta_{d}}\right)^{1/2}\approx0.054$. Finally, we find for the ``good" triple coincidence rate $R_{T,\mathrm{good}}\approx\frac{1}{2} \eta_{d}^3 \left|\gamma_{1}\right|^2\left|\gamma_{2,3}\right|^2\approx0.12\,\si\hertz$. The corresponding experimentally observed rate was $R_{T}=(0.16\pm0.03)\,\si{\hertz}$ as mentioned in the main text.

\end{document}